\documentclass[final,5p,times]{elsarticle}
\usepackage{amsmath}
\usepackage{amssymb}
\usepackage{amsfonts}
\usepackage{graphicx}
\usepackage{subfig}
\usepackage{threeparttable}
\usepackage{bm}
\usepackage{xcolor}
\usepackage{latexsym} 
\usepackage{enumerate}
\usepackage{bm}
\usepackage{ulem}
\usepackage[colorlinks,
linkcolor=blue,
anchorcolor=blue,
urlcolor=red,
citecolor=blue]{hyperref}
\biboptions{sort&compress}
\journal{Physics Letters B}

\begin{document}
\begin{frontmatter}

\title{Unlocking the initial neutron density distribution from the two-pion HBT correlation function in heavy-ion collisions}

\author[1]{Pengcheng Li}
\author[1,3]{Manzi Nan}
\author[2]{Haojie Zhang}
\author[2]{Junhuai Xu}
\author[1]{Xilong Xiang}
\author[2]{Yijie Wang}
\author[1]{Yongjia Wang}
\author[3]{Gaochan Yong}
\author[4]{Tadaaki Isobe}
\author[2]{Zhigang Xiao\corref{cor1}}
\author[1,3]{Qingfeng Li\corref{cor2}}

\cortext[cor1]{Corresponding author:~xiaozg@tsinghua.edu.cn}
\cortext[cor2]{Corresponding author:~liqf@zjhu.edu.cn}

\address[1]{School of Science, Huzhou University, Huzhou 313000, China}
\address[2]{Department of Physics, Tsinghua University, Beijing 100084, China}
\address[3]{Institute of Modern Physics, Chinese Academy of Sciences, Lanzhou 730000, China}
\address[4]{RIKEN Nishina Center, Hirosawa 2-1, Wako, Saitama 351-0198, Japan}

\begin{abstract}
Revealing the neutron density distribution in the nucleus is one of the crucial tasks of nuclear physics. 
Within the framework of the ultrarelativistic quantum molecular dynamic model followed by a correlation afterburner program, we investigate the effects of the initial neutron density distribution on the charged-pion yield ratio $\pi^{-}/\pi^{+}$, the two-pion momentum correlation function, and the emission source dimension. 
It is found that the $\pi^{-}/\pi^{+}$ ratio is sensitive to the initial neutron density distribution and the impact parameter, especially for collisions at large impact parameter.
However, the charge splitting in the correlation functions between positively $\pi^{+}\pi^{+}$ and negatively $\pi^{-}\pi^{-}$, as well as the source radii and volumes extracted exhibit a stronger dependence on the initial neutron density distribution, but a weaker dependence on the impact parameter.
The present study highlights that  $\pi^{+}\pi^{+}$ and $\pi^{-}\pi^{-}$ correlation functions in heavy-ion collisions could be used to probe the initial neutron density distribution of nuclei.
\end{abstract}


\end{frontmatter}

\section{Introduction}\label{section1}
The distributions of proton and neutron in neutron-rich nuclei are the fundamental physical quantities in nuclear physics, and have attracted much attention in nuclear physics and astrophysics \cite{Tsang:2012se,Fattoyev:2017jql}. 
The proton distributions in nuclei can be determined with very high accuracy via electron scattering experiments \cite{DeVries:1987atn}, however, the knowledge of neutron distribution in nuclei is still relatively limited due to the understanding of strong interactions in nuclei being inconsistent yet. 
In neutron-rich nuclei, the excess neutrons tend to be on the surface of the nuclei, forming the neutron skin or neutron halo \cite{Thiel:2019tkm,PREX:2021umo}. 
The neutron skin thickness is commonly defined as the difference between the neutron and the proton root-mean-square (RMS) radii: $\Delta R_{np}=\langle r^{2}\rangle_{n}^{1/2}-\langle r^{2}\rangle_{p}^{1/2}$. 
However, one cannot infer more reliable information on neutron distribution in the neutron-rich nuclei from the $\Delta R_{np}$ \cite{Trzcinska:2001sy,Pawlowski:2004uc}. 

The heavy-ion collisions (HICs) have been widely used to constrain the neutron distributions, the neutron skin thickness and image shapes of atomic nuclei \cite{Pawlowski:2004uc,Li:2019kkh,Giacalone:2023cet,STAR:2024wgy,Xu:2024bdh}.  
Based on numerous studies using theoretical models, strong correlations have been found between the neutron skin thickness and the final state observables, such as the charged-pion ratio ($\pi^{-}/\pi^{+}$) \cite{Wei:2013sfa,Hartnack:2018sih,Yang:2020wue}, neutron/proton ratio ($n/p$) \cite{Sun:2009wf}, $^{3}\rm{H}/^{3}\rm{He}$ ratio \cite{Dai:2014rja} and the difference of momentum between neutrons and protons \cite{Ding:2024jfr}, in nuclear reactions induced by neutron-rich nuclei \cite{Ding:2024xxu}. 
In addition, by directly analyzing the final state charged hadron multiplicity and eccentricity in isobar collisions, it was found that these final state observables are exquisitely sensitive to the neutron skin type and thickness \cite{Li:2019kkh,Xu:2021vpn}. 
And by analyzing the collective flows of particles emerging from the emission source, and comparing them with models of hydrodynamic expansion for different quark-gluon plasma (QGP) shapes, the shapes of the originally colliding nuclei can be inferred \cite{STAR:2024wgy}. 

The correlation function method has been applied to study the spatial-temporal distribution of the particle emission source and the final state interactions between particle pairs over a wide energy range \cite{Verde:2006dh,Wang:2021mrv,FOPI:2004wcp,Si:2025eou,STAR:2024zvj,Wang:2018avn,Wang:2022nxr,Li:2012ta,Li:2010ew}. 
For the nucleon-nucleon pair case, the $N\text-N$, momentum correlation function has been suggested to explore the high momentum tail of the nucleon momentum distribution \cite{Wei:2019jva} and the nuclear symmetry energy \cite{Chen:2002ym,Li:2009kxa}. In Ref. \cite{Ma:2006ck}, it was pointed out that the $N\text- N$ correlation function carries the information of the neutron- or proton-halo in exotic nuclei. Interestingly, the density distributions of the valence neutrons will significantly influence the $N\text- N$  correlation function at large impact parameters and high incident energies \cite{Cao:2012cv}. 
In addition, the non-monotonic energy dependence of the two-pion Hanbury-Brown-Twiss (HBT) radii parameters extracted from the correlation function serves as a sensitive signal of the phase transition from hadron gas to QGP \cite{Pratt:1984su,Pratt:1986cc}, since they are sensitive to the dense nuclear equation-of-state (EoS), and have been widely investigated and measured \cite{E895:2000opr,STAR:2001gzb,Lacey:2014wqa,Zhang:2017axr,Batyuk:2017smw,HADES:2019lek,Li:2022icu,Li:2022iil,Li:2008qm,Li:2007yd}. 

Recently, an obvious collision energy-dependent splitting in the HBT radii (source radii) between negatively charged-pion pairs $\pi^{-}\pi^{-}$ and positively charged-pion pairs $\pi^{+}\pi^{+}$, especially at low transverse momentum, was observed by the HADES and STAR Collaborations in central Au+Au collisions at GeV energies, the sideward ($R_{S}$), outward ($R_{O}$) and longitudinal ($R_{L}$) radii of $\pi^{-}\pi^{-}$ are larger than those of $\pi^{+}\pi^{+}$~\cite{HADES:2018gop,Luong:2024eaq}. 
However, the explanations for this splitting are insufficient and have not reached a consensus. 
Within the isospin-dependent quantum molecular dynamics model simulation, it was found that these splittings are caused by the two-body Coulomb interaction~\cite{Fang:2022kru}, in accordance with an earlier study in~\cite{Zajc:1984vb}. However, the STAR preliminary results show that these splittings can be largely described by the third-body Coulomb effect~\cite{QM25_VinhLuong}. 

In our previous works~\cite{Li:2022icu,Li:2022iil}, within the framework of the ultra-relativistic quantum molecular dynamics (UrQMD) model, the effects of resonance decay widths, in-medium nucleon-nucleon (in)elastic cross sections, and potentials on the two-pion interferometry are investigated. 
The results suggest that the Coulomb interaction is the main cause of the charge splitting of the source radii, but other factors, such as the initial neutron and proton density distributions and the treatment of the dynamic process of HICs, may also contribute to this charge splitting. 
It deserves further investigation, since the neutron skin thickness can be directly inferred from the initial neutron and proton density distributions. In the present work, we are motivated to study the effect of the initial proton and neutron profile on the charged pion yield and ratio, as well as on the HBT correlation. The neutron density distributions have been taken into account in the UrQMD model in the initiation. 
The paper is organized as follows: In Sec. \ref{sec:2}, the UrQMD model and methods are briefly described.
In Sec. \ref{sec:3}, the effects of the initial neutron density distribution on the observables are shown and discussed. 
Finally, the summary and outlook are presented in Sec. \ref{sec:4}.

\section{Framework description}\label{sec:2}

In this work, the UrQMD model~\cite{Bass:1998ca,Bleicher:1999xi} is adopted for simulations, and two modes can be used to initialize the colliding nuclei in the model.  
One is the hard sphere mode, where the initial nuclei are prepared by sampling the centroids of wave packets $r_{i}$ in a hard sphere with radius $R$. 
Another is the Woods-Saxon mode, where the positions of the nucleons are sampled within the Woods-Saxon form, 
\begin{equation}
\rho(r)=\frac{\rho_{0}}{1+\exp{(\frac{r-R}{a}})}, 
\end{equation}
where $a=0.545$ fm is the diffuseness parameter. 
To explore the effects of the initial neutron density distribution on the final state observables, the different density distributions of protons and neutrons from the Droplet model are used here~\cite{Sun:2009wf}. 
\begin{equation}\label{fiti}
\rho_{i}(r)=\frac{\rho_{i}^{0}}{1+\exp{\left( \frac{r-R_{i}[1-(0.413f_{i}t_{i}/R_{i})^2]}{f_{i}t_{i}/4.4} \right) }},~~i=n,p,
\end{equation}
here, $\rho_{i}^{0}$ is the normalization constant, $t_{i}=2.18$ is the diffuseness parameter and can be changed by adjusting the $f_{i}$. 
For neutron-rich nuclei, $f_{p}=1.0$ is adopted as in the Droplet model, while different neutron density distributions and neutron skin thickness $\Delta R_{np}$ can be obtained by adjusting $f_{n}$. 
Within this approach, the effects of the neutron skin thickness on the yield ratio of neutrons to protons, tritons to $^{3}\rm{He}$ have been investigated~\cite{Sun:2009wf,Dai:2014rja,Ding:2024jfr,Ding:2024xxu}.

Then, the stable initialized nuclei will be selected with strict conditions and propagated within the transport model. 
In this work, the mean-field mode of the UrQMD model is used to simulate Au+Au collisions at $\sqrt{s_{NN}}=3$ GeV, with the evolution stopped at 100 fm/c \cite{Hillmann:2018nmd,Hillmann:2019wlt,Li:2022iil,Steinheimer:2022gqb}. 
In this mode, the potential energy includes the two-body and three-body Skyrme interactions, as well as the two-body Coulomb interaction,
\begin{equation}\label{Sky_Eos}
V=V_{\rm{Sky.}}^{(2)}+V_{\rm{Sky.}}^{(3)}+V_{\rm{Coul.}}.
\end{equation}
Recently, Ref.~\cite{Li:2025iqq} has highlighted that the nuclear EoS will gradually harden as density and temperature increase in HICs with higher beam energies. 
Moreover, the stiffness of dense nuclear matter comes partly from the repulsive single-particle potential due to its momentum dependence. 
Thus, a hard EoS ($K_{0}$=380 MeV) without momentum-dependence is used here, and it has been shown to provide a good description of available data for HICs at GeV energies~\cite{Hillmann:2018nmd,Hillmann:2019wlt,Li:2022iil,Steinheimer:2022gqb,STAR:2021yiu}.

After the UrQMD model simulations, the space-time coordinates and four-momentum of particles after the last interaction (collisions or decay) serve as the input for the standard correlation after-burner (CRAB) code to construct the HBT correlation function $C$~\cite{Lisa:2005dd,Pratt:2008sz,Li:2008bk}. 
Lastly, the correlation function is fitted assuming a three-dimensional Gaussian form in the longitudinally comoving system, resulting in the HBT radii of the particle emission source~\cite{Li:2008bk,Li:2008ge}. 
In this work, the Coulomb final-state interactions are also taken into account in the CRAB program and fitting process. And this method is well established in the larger energy regime. For a more systematic description, we refer the reader to~\cite{Lisa:2005dd,Li:2022iil,Li:2022icu,Pratt:2008sz,Li:2008bk,Li:2008ge}.

\section{Results and discussions}\label{sec:3}

\subsection{Density profile and evolution}\label{sec:31}
\begin{figure}[t!]
\centering
\includegraphics[width=0.5\textwidth]{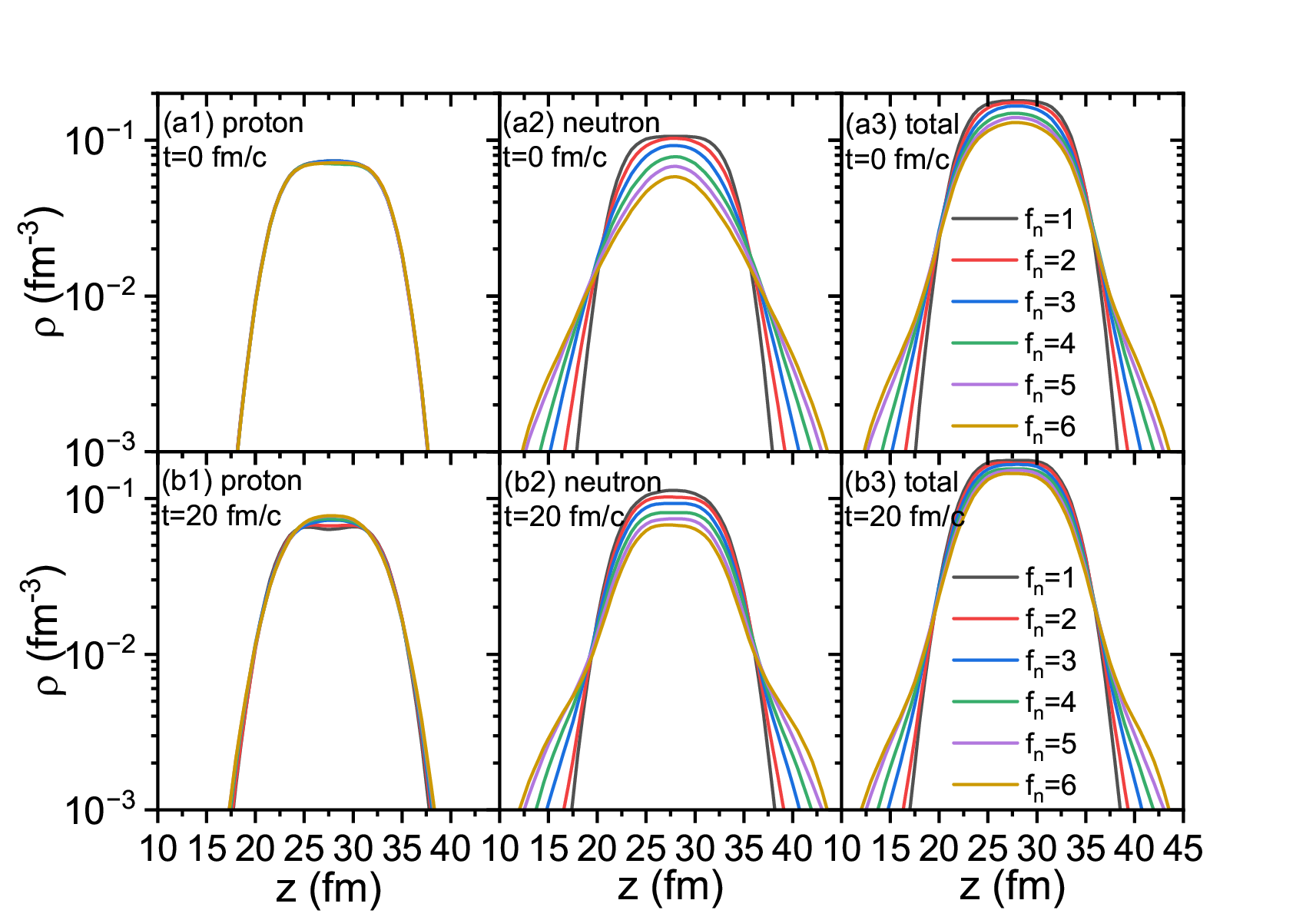}
\caption {\label{fig1}(Color online) The proton (left), neutron (middle), and total (right) density distributions of the target nucleus $^{197}\rm{Au}$ calculated with $f_{n}$ varies from 1 to 6 in Eq.~\ref{fiti}. The top panels display the initial density profiles, while the bottom panels present the density profiles at $t=20$ fm/c.}
\end{figure}

To check the stability of the initial configurations as described in the previous section, the time evolution of radial density distributions of colliding nuclei is analyzed. 
The successfully sampled target and projectile are picked out from the model simulation, the initial ($t=0$ fm/c) density profiles and the density profiles at $t=20$ fm/c of target nucleus $^{197}\rm{Au}$ are shown in Fig.~\ref{fig1}(a1-a3) and Fig.~\ref{fig1}(b1-b3), respectively. Here, the distance between the target and projectile nuclei at initialization is set to 20 fm. 
At $t=0$ fm/c, the width of the proton density distribution is nearly identical across all calculations, since $f_{p}=1$ remains unchanged. 
In contrast, the difference between neutron and proton distributions increases with $f_n$ varying from 1 to 6.  
Notably, the density profiles remain stable within the simulation time of 20 fm/c, indicating that the initialized nuclei maintain their structural integrity. For the investigated Au+Au collisions at $\sqrt{s_{NN}}=3$ GeV, this duration is sufficient to achieve full nuclear overlap \cite{Reichert:2024ayg}. 
Then, the question arises: how to characterise the difference in the initial neutron and proton density distribution by using some of the final-state observables?

\subsection{Charged-pion yields and ratios}

\begin{figure}[t!]
\centering
\includegraphics[width=0.49\textwidth]{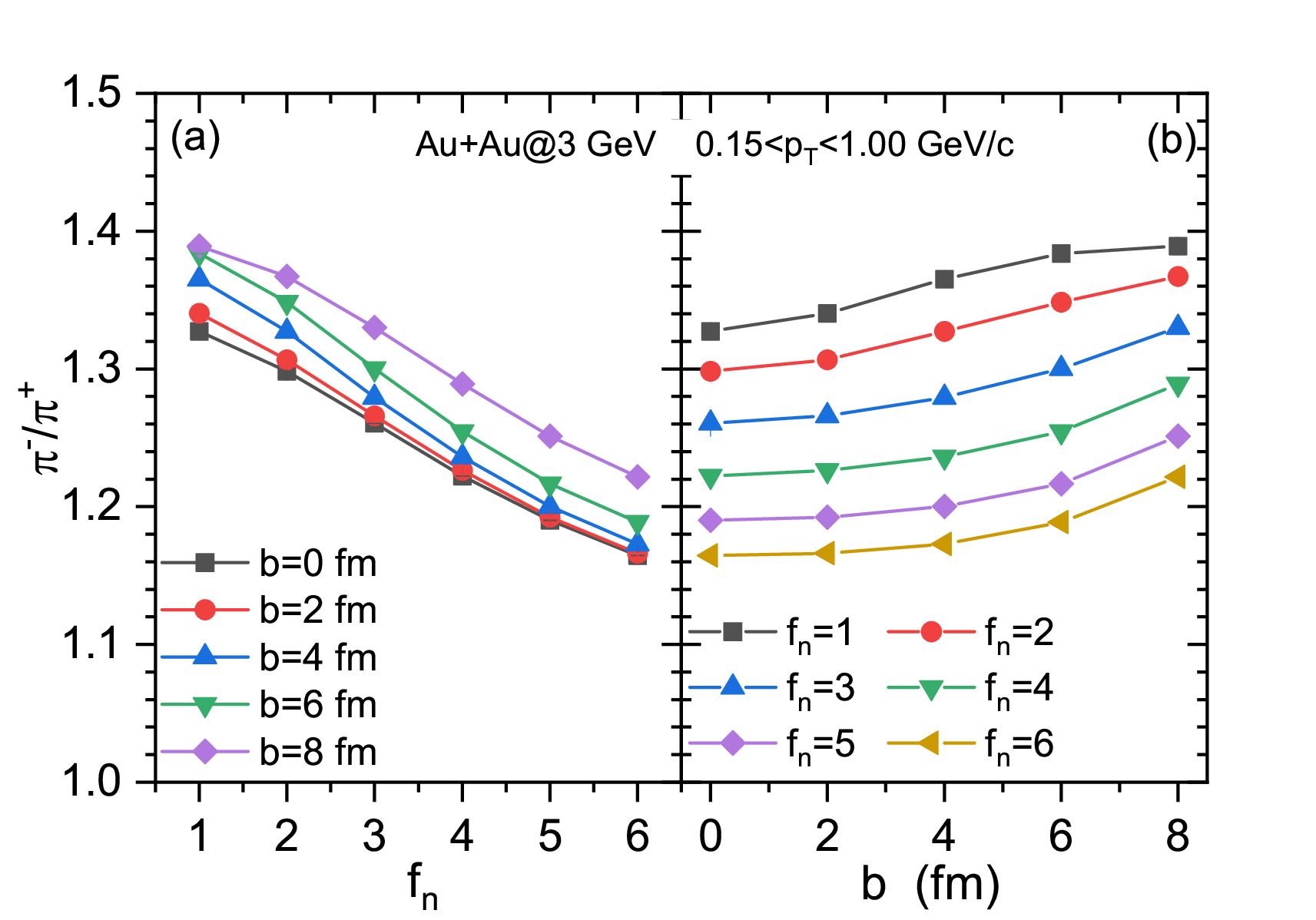}
\caption {\label{fig2}(Color online) The dependence of $\pi^{-}/\pi^{+}$ on the diffuseness parameter of neutron density distribution $f_{n}$ [panel (a)] and the impact parameter $b$ [panel (b)] under the condition of $0.15<p_{T}<1.0$ GeV/c for Au+Au at $\sqrt{s_{NN}}=3$ GeV.}
\end{figure}

To answer the above question, we first examine the charged-pion ratio at the final state. 
Fig.~\ref{fig2} presents the $\pi^{-}/\pi^{+}$ ratio as a function of the diffuseness parameter of neutron density distribution $f_{n}$ [panel (a)] and the impact parameter $b$ [panel (b)]. 
A clear and monotonically decreasing trend is observed in Fig.~\ref{fig2}(a), where the ratio decreases with increasing $f_{n}$, indicating that the $\pi^{-}/\pi^{+}$ ratio is sensitive to the initial neutron density distribution of the colliding nuclei. 
Moreover, the $\pi^{-}/\pi^{+}$ ratios for $b = 0, 2, 4$ fm are relatively close to each other, especially at larger $f_{n}$, whereas those for $b = 4$, 6, and 8 fm show more distinct separations. 
To better illustrate the dependence of the $\pi^{-}/\pi^{+}$ ratio on $b$, Fig.~\ref{fig2}(b) exhibits the ratio as a function of the $b$ for different values of $f_{n}$. 
In central collisions ($b=0$, 2 fm), the charged-pion ratio shows a slight increase with increasing $b$ across all six cases ($f_{n}=1\sim6$), while in peripheral collisions ($b=4$, 6, and 8 fm), the ratio increases more rapidly with $b$.  
These conclusions are consistent with Ref.~\cite{Wei:2013sfa}, indicating that in peripheral collisions, the charged-pion ratio is strongly influenced by both the neutron density distribution of the colliding nuclei and the impact parameter.

\begin{figure}[b!]
\centering
\includegraphics[width=0.49\textwidth]{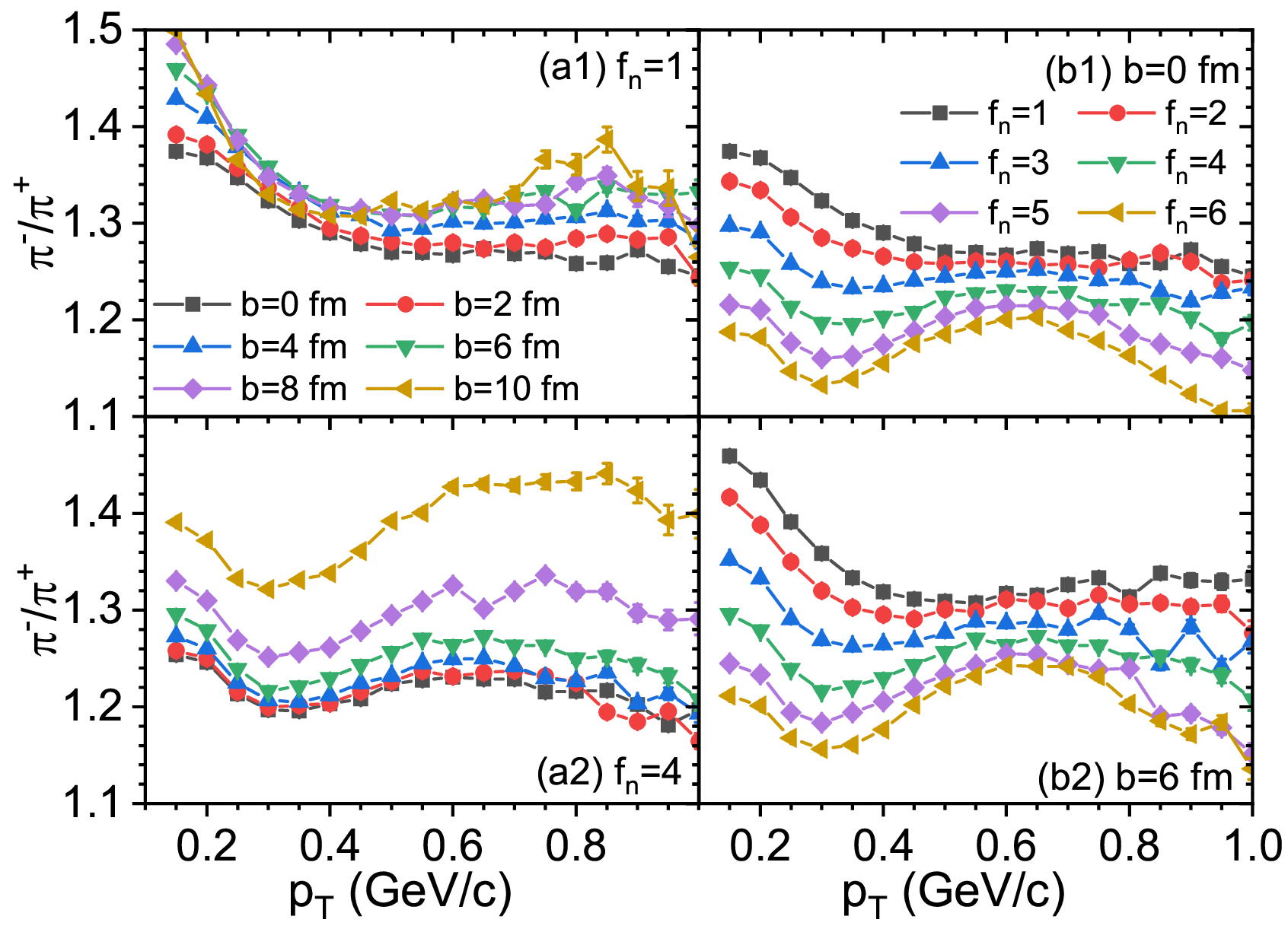}
\caption {\label{fig3}(Color online) The dependence of $\pi^{-}/\pi^{+}$ ratio on transverse momentum $p_{T}$ in Au+Au collisions at $\sqrt{s_{NN}}=3$ GeV for $0.15<p_{T}<1.0$ GeV/c. Results from simulations with varying $b$ (left panels) and $f_{n}$ (right panels) are represented by different symbols.}
\end{figure}

\begin{figure*}[h!]
\centering
\includegraphics[width=0.85\textwidth]{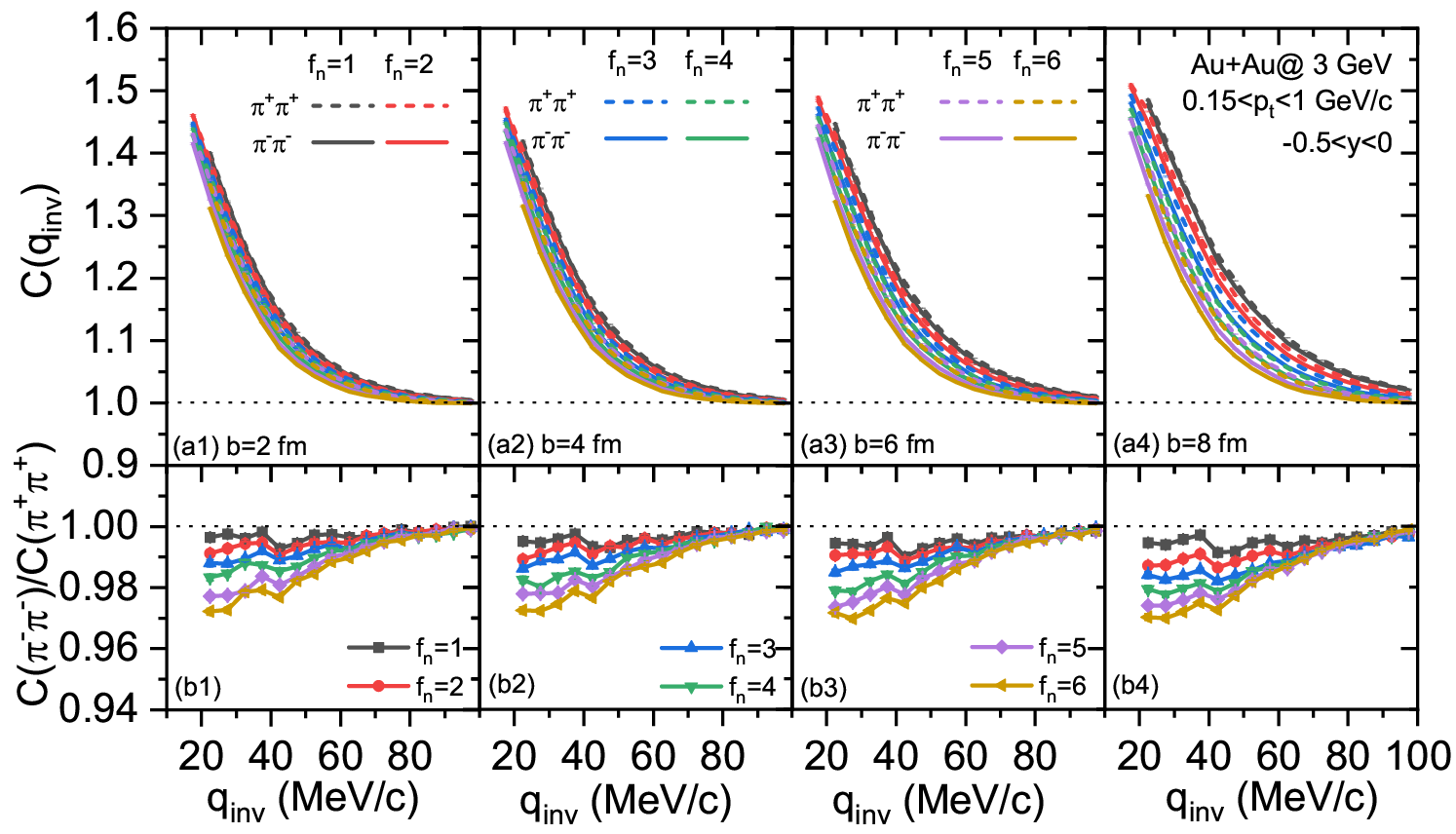}
\caption {\label{fig4}(Color online) Top panels: the one-dimensional $\pi^{+}\pi^{+}$ and $\pi^{-}\pi^{-}$ correlation functions as a function of the invariant relative momentum $q_{\rm{inv}}$ in Au+Au at $\sqrt{s_{NN}}=3$ GeV with various $b$ and different $f_{n}$. 
Bottom panels: the corresponding correlation function ratios between $\pi^{-}\pi^{-}$ and $\pi^{+}\pi^{+}$. 
The horizontal dotted lines represent unity.
}
\end{figure*}

In order to further understand the dependence of the $\pi^{-}/\pi^{+}$ ratio on $f_{n}$ and $b$, simulations with varying $f_{n}$ values are presented as a function of transverse momentum $p_{T}$ in Fig.~\ref{fig3}. 
For $f_{n}=1$ [panel (a1)], the ratio decreases with increasing $p_{T}$ and eventually saturates, showing only a weak dependence on $b$. 
In contrast, for larger $f_{n}$ values [panel (a2)], the ratio displays a pronounced hump at mid-$p_{T}$, and the $b$ dependence becomes more significant, particularly in peripheral collisions.  
Furthermore, as $f_{n}$ increases [panels (b1) and (b2)], the $\pi^{-}/\pi^{+}$ ratio decreases across the entire $p_{T}$ range. 
Notably, the $\pi^{-}/\pi^{+}$ ratios diverge significantly at low $p_{T}$ for different $f_{n}$, consistent with the findings in Ref.~\cite{Ding:2024jfr}. And the mid-$p_{T}$ hump in the $\pi^{-}/\pi^{+}$ ratio also becomes more pronounced with increasing $f_{n}$. 
As shown in Fig.~\ref{fig1}(a2) and (b2), increasing in $f_{n}$ leads to a reduced neutron density in the nuclear interior and an enhanced neutron density at the surface, while the proton density remains largely unchanged throughout. Consequently, $\pi^{-}$ with low-$p_{T}$, which originate predominantly from the dense inner region, become less abundant, leading to a decrease in the $\pi^{-}/\pi^{+}$ ratio at low $p_{T}$. In contrast, $\pi^{-}$ with high-$p_T$, which emerge mainly from the neutron-rich surface (i.e., the skin–skin interaction region), become more pronounced, giving rise to the observed hump in the $\pi^{-}/\pi^{+}$ ratio. 

\subsection{HBT correlation functions of charged-pion pairs}

\begin{figure}[b!]
\centering
\includegraphics[width=0.49\textwidth]{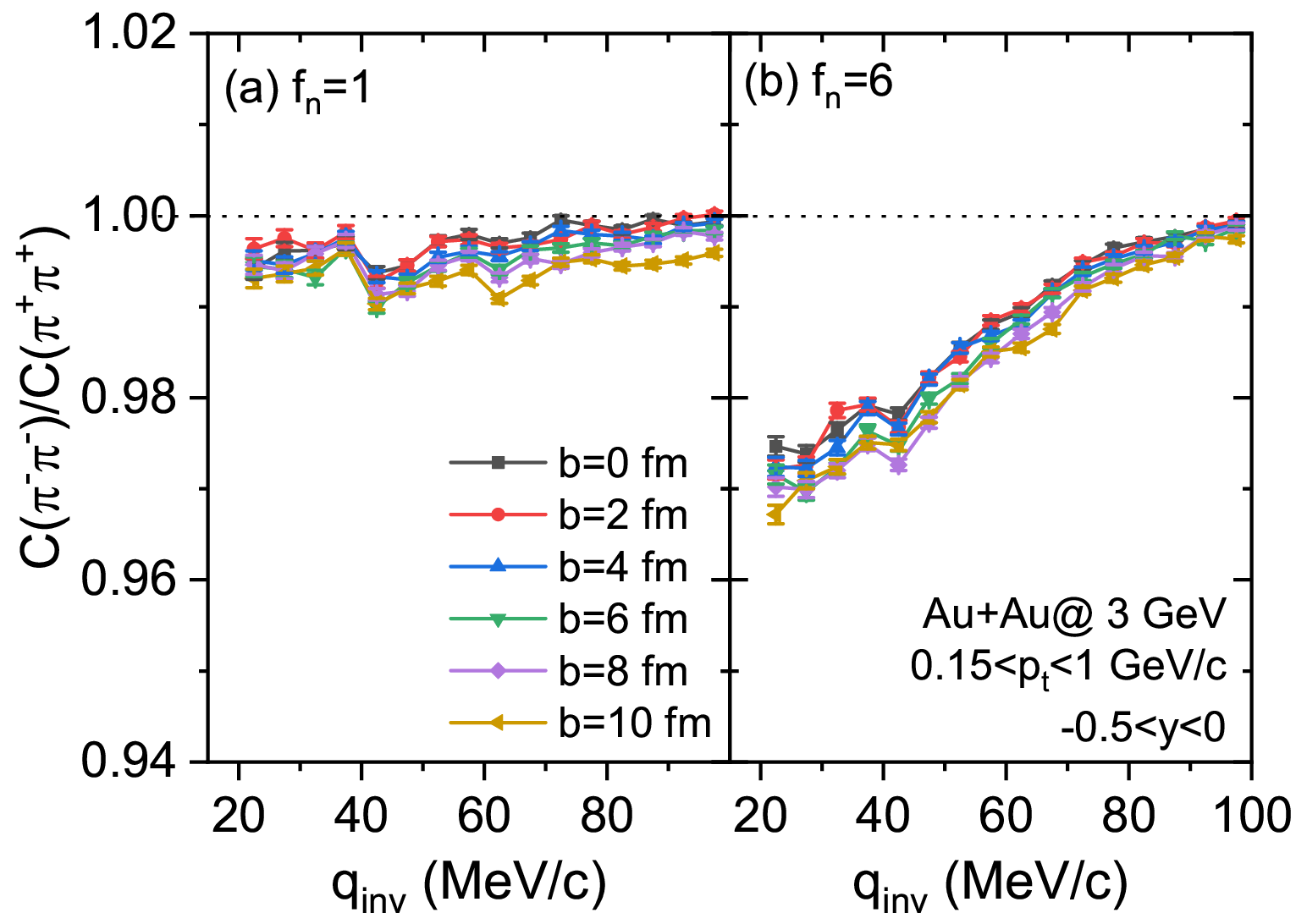}
\caption {\label{fig5}(Color online) The HBT correlation function ratios between $\pi^{-}\pi^{-}$ and $\pi^{+}\pi^{+}$ for $f_{n}=1$ (left panel) and $f_{n}=6$ (right panel) with various $b$. }
\end{figure}

Then, let us further turn to the HBT correlation functions of charged-pion pairs. 
Figure~\ref{fig4}(a1–a4) presents the one-dimensional correlation functions for $\pi^{+}\pi^{+}$ (dashed lines) and $\pi^{-}\pi^{-}$ (solid lines) as a function of the invariant relative momentum $q_{\rm{inv}}$, calculated using the CRAB program. 
Simulations are performed for $b$ = 2, 4, 6, and 8 fm, with the $f_n$ varied from 1 to 6.
Charged pions are selected within the transverse momentum range $0.15 < p_{T} < 1.0$ GeV/c and rapidity window $-0.5 < y < 0$, following the STAR experimental conditions~\cite{Luong:2024eaq}. 
The correlation functions exhibit a clear dependence on the initial neutron density distribution, particularly at larger impact parameters. 
This effect is most pronounced at $b = 8$ fm, where the $\pi^{-}\pi^{-}$ and $\pi^{+}\pi^{+}$ correlation functions show the largest separation across different $f_n$ values. 
And as $b$ decreases, this separation becomes less distinct. 
The corresponding ratios of $\pi^{-}\pi^{-}$ to $\pi^{+}\pi^{+}$ correlation functions are displayed in Fig.~\ref{fig4}(b1–b4).  
It can be found that for all four impact parameter sets, the ratio is close to unity when $f_{n}=1$, but systematically decreases as $f_{n}$ increases. 
This trend indicates that a greater disparity in the initial density distributions of protons and neutrons leads to an enhanced difference in the correlation functions of $\pi^{-}\pi^{-}$ and $\pi^{+}\pi^{+}$. 
Notably, this behavior is similar to $f_{n}$ dependence of the $\pi^{-}/\pi^{+}$ yield ratio presented in Figs.~\ref{fig2} and \ref{fig3}.

Figure~\ref{fig5} shows the correlation function ratios for fixed values of $f_{n}$ (especially, $f_n=1$ in the left panel and $f_n=6$ in the right panel) across different $b$. 
Within statistical uncertainties, the ratios remain almost unchanged over the full range of $b$, including for peripheral collisions at even $b=$~10 fm. 
This insensitivity to $b$ arises because, while changing $b$ alters the geometry of the overlap region in the collision, it does not affect the intrinsic initial density difference between protons and neutrons for a fixed $f_n$. As a result, although the individual correlation functions for $\pi^{+}\pi^{+}$ and $\pi^{-}\pi^{-}$ are modified by $b$, the difference in correlation functions between $\pi^{+}\pi^{+}$ and $\pi^{-}\pi^{-}$ and the corresponding ratio exhibit only weak dependence on $b$. 
These results indicate that the observed difference between $\pi^{+}\pi^{+}$ and $\pi^{-}\pi^{-}$ correlation functions is rather sensitive to the initial density difference between protons and neutrons, but only minor sensitivity to the collision centrality. 

\begin{figure}[t!]
\centering
\includegraphics[width=0.5\textwidth]{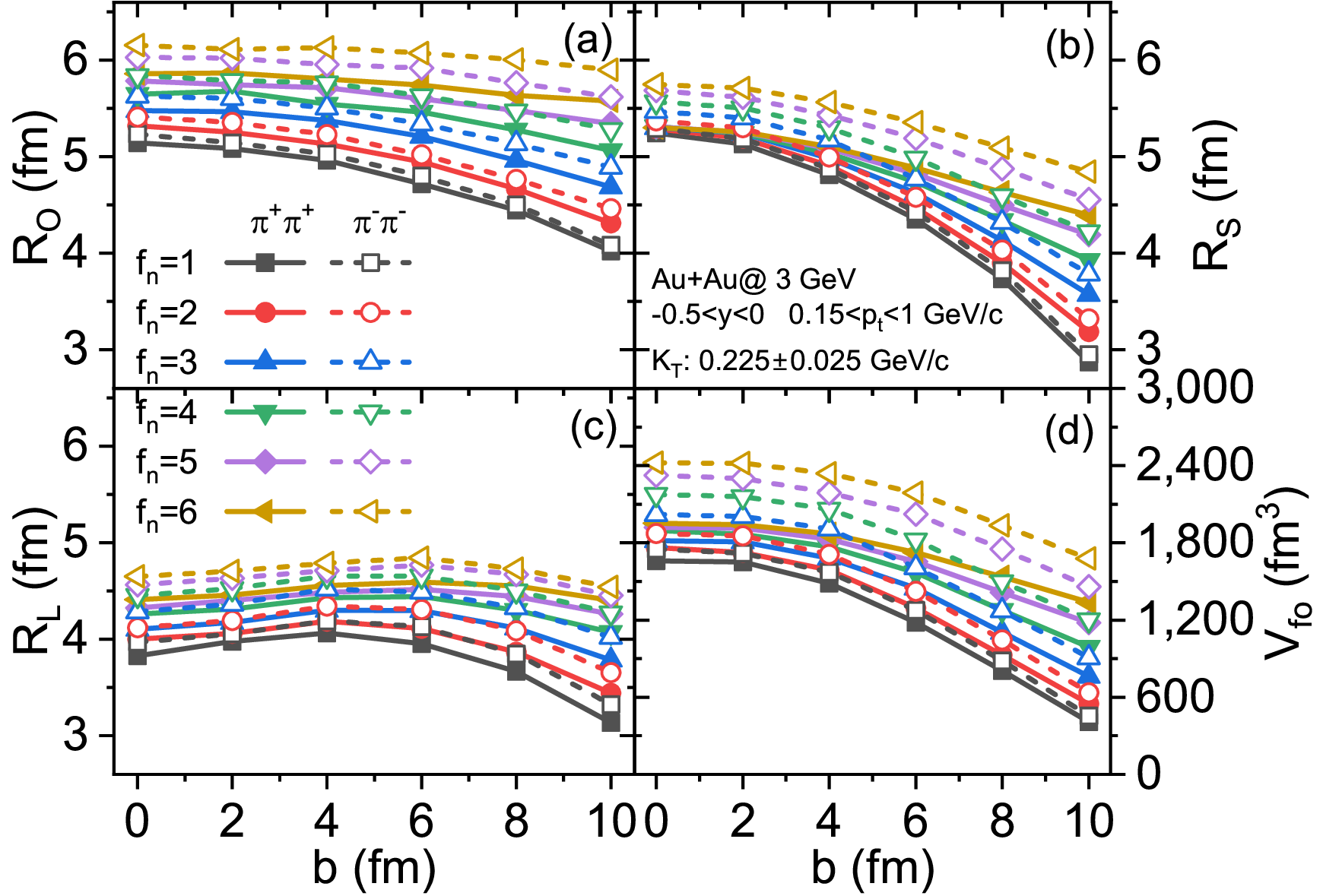}
\caption {\label{fig6}(Color online) The $b$ dependence of the $R_{L}$, $R_{O}$, $R_{S}$ and $V_{\rm{fo}}$ extracted from the $\pi^{+}\pi^{+}$ and $\pi^{-}\pi^{-}$ correlation functions. Solid symbols and lines display the results from $\pi^{+}\pi^{+}$ correlation functions, while the open symbols and dashed lines represent the results from $\pi^{-}\pi^{-}$ correlation functions. 
}
\end{figure}

\begin{figure}[t!]
\centering
\includegraphics[width=0.49\textwidth]{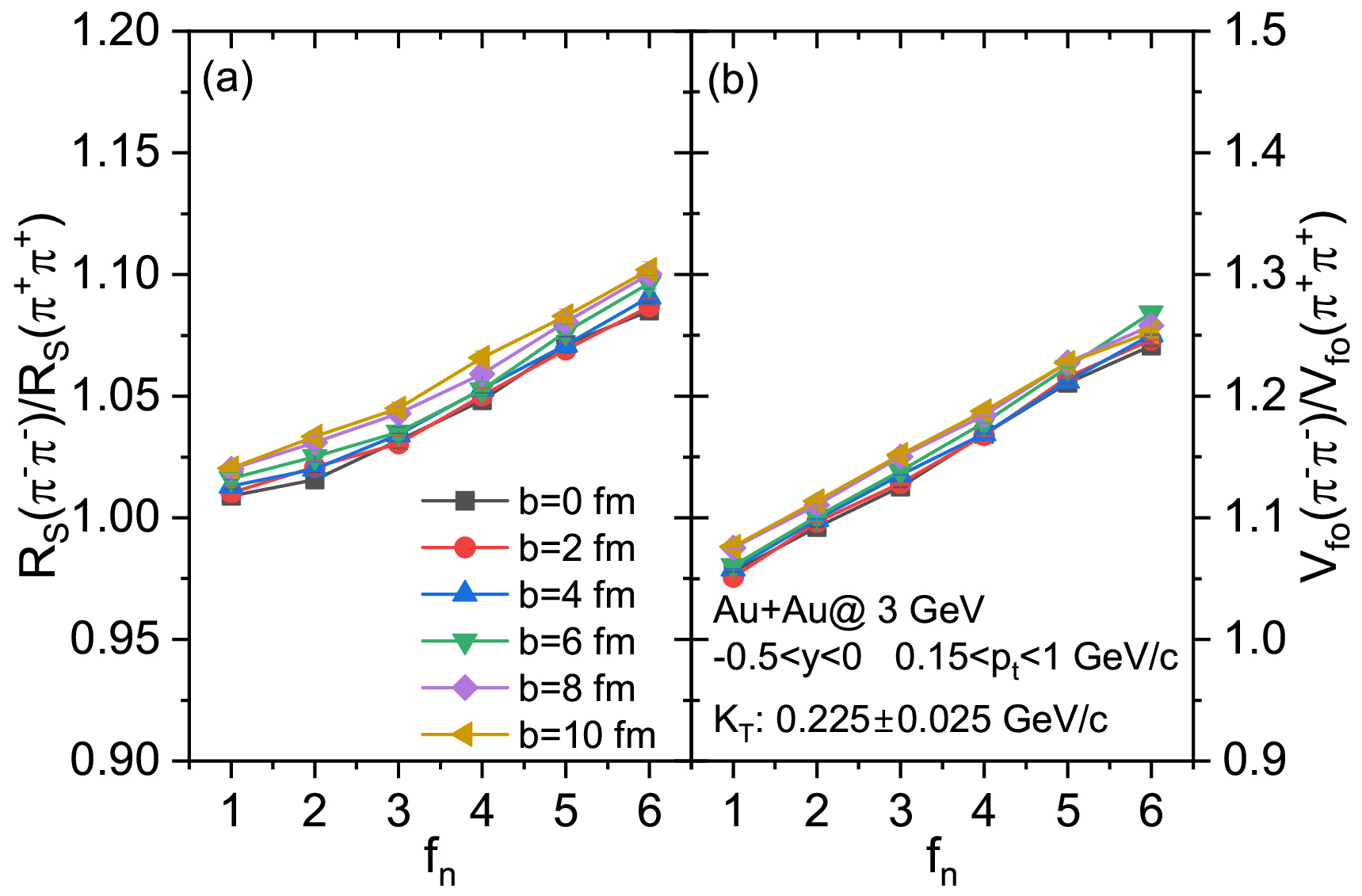}
\caption {\label{fig7}(Color online) The $R_{S}$ (left panel) and $V_{\rm{fo}}$ (right panel) ratio between $\pi^{-}\pi^{-}$ and $\pi^{+}\pi^{+}$ as a function of $f_{n}$. 
}
\end{figure}

Finally, by fitting the correlation functions with assuming a three-dimensional Gaussian form in the longitudinally comoving system, one can obtain the three-dimensional HBT radii of the particle emission source, $R_{L}$, $R_{O}$, and $R_{S}$, corresponding to the longitudinal (i.e., the beam direction), the outward (the direction of the transverse component of the pair-relative momentum $\textbf{k}_{T}=(\textbf{p}_{T,1}+\textbf{p}_{T,2})/2$), and the sideward directions (perpendicular to the other two directions), respectively \cite{Li:2022iil,HADES:2019lek}. 

Figure~\ref{fig6} shows the impact parameter dependence of the $R_{O}$, $R_{S}$, $R_{L}$ and the corresponding pion emission source volume, $V_{\rm{fo}}=(2\pi)^{3/2}R_{L}R_{S}^{2}$, extracted from $\pi^{-}\pi^{-}$ (open symbols) and $\pi^{+}\pi^{+}$ (solid symbols) correlation functions. 
As $b$ increases from 0 to 10~fm, both the radii and the source volume systematically decrease, reflecting the reduced geometric overlap of the colliding nuclei. 
In addition, the $R_{O}$ and $R_{L}$ exhibit a weaker dependence on $b$, while the $R_{S}$ and hence $V_{\rm{fo}}$ show a more obvious $b$ dependence. 
However, the difference between $\pi^{-}\pi^{-}$ and $\pi^{+}\pi^{+}$ results for each value of $f_{n}$ remain nearly constant across the full range of $b$. 

For more clearly, Fig.~\ref{fig7} further depicts the $f_{n}$ dependence of the $R_{S}$ and $V_{\rm{fo}}$ ratios between $\pi^{-}$ and $\pi^{+}$ emission source for different $b$. 
These ratios demonstrate a clear sensitivity to $f_n$ but a weak dependence on $b$. 
As $f_{n}$ increases, the initial neutron distribution radii and the difference in the initial density distribution between protons and neutrons become larger, resulting in the size of $\pi^{-}$ emission source increases, and the charge splittings in these physical quantities between $\pi^{+}\pi^{+}$ and $\pi^{-}\pi^{-}$ are enlarged. 
These results reinforce the conclusion that the charge splitting in the two-pion HBT correlation functions (as shown in Fig.~\ref{fig5}) and the HBT radii parameters are much more sensitive to $f_n$ rather than the collision centrality. 

\begin{figure}[t!]
\centering
\includegraphics[width=0.45\textwidth]{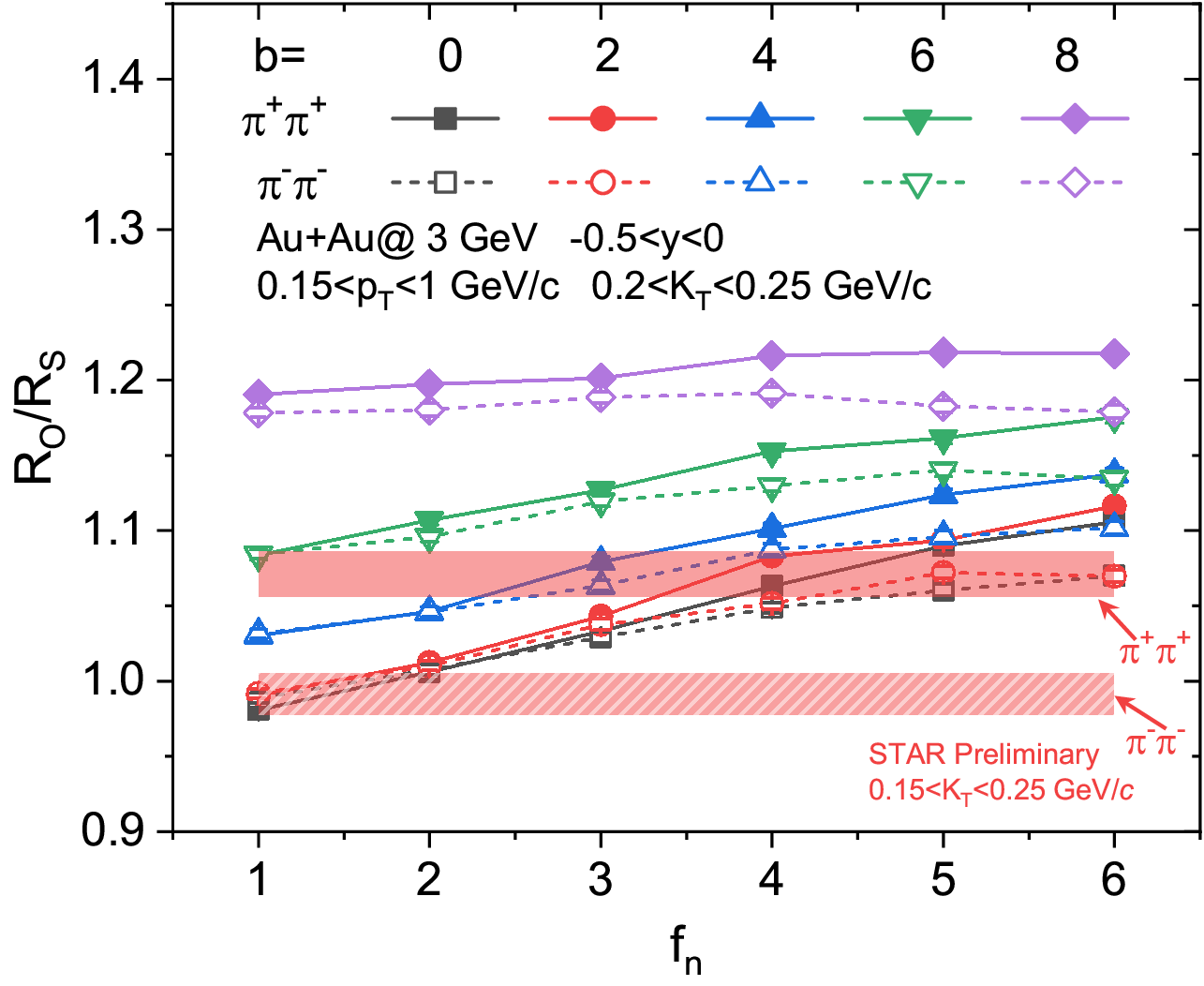}
\caption {\label{fig8}(Color online) The $f_{n}$ dependence of the $R_{O}/R_{S}$ extracted from the $\pi^{+}\pi^{+}$ and $\pi^{-}\pi^{-}$ correlation functions. The STAR Preliminary data for 0-10\% Au+Au collisions with $0.15<k_{T}<0.25$ GeV/c are taken from Ref. \cite{Luong:2024eaq}.}
\end{figure}

Additionally, Fig. \ref {fig8} shows the $R_{O}/R_{S}$ as a function of the $f_{n}$, in which the $R_{O}/R_{S}$ ratio for $\pi^{+}\pi^{+}$ (solid symbols) is observed to be larger than that for $\pi^{-}\pi^{-}$ (open symbols), which is consistent with the STAR preliminary data \cite{Luong:2024eaq}. 
And this available experimental data on the HBT radii parameter $R_{O}/R_{S}$ favors $f_{n}<3$ for $^{197}$Au. 
Moreover, similar to the HBT correlation function ratios, the HBT radii ratio and the emission source volume $V_{\rm{fo}}$ ratio between $\pi^{-}\pi^{-}$ and $\pi^{+}\pi^{+}$ as presented in Figs. \ref{fig4} to \ref{fig7}, exhibit a clear charge splitting. 
This splitting is also evident in the $R_O/R_S$ ratio, again demonstrating a pronounced sensitivity to $f_{n}$, while showing relatively weak dependence on $b$. 
In contrast, the charged-pion yield ratio $\pi^{-}/\pi^{+}$, as shown in Figs. \ref{fig2} and \ref{fig3}, depends on both $f_{n}$ and $b$, with a particularly pronounced $b$ dependence in peripheral collisions. 
This comparison suggests that, unlike yield observables, the momentum-space correlations of two identical particles (such as HBT correlations) can effectively suppress the influence of the impact parameter. Consequently, two-pion HBT correlation offers a more robust means of constraining the initial neutron-proton density distribution in HICs.

\section{Summary and outlook}\label{sec:4}

In summary, by varying the diffuseness parameter $f_{n}$ of the neutron density distribution, a difference is introduced between the neutron and proton density profiles. Then the effects of the initial neutron density distribution on the charged-pion yield ratio $\pi^{-}/\pi^{+}$, momentum correlation functions, and HBT radii parameters in Au+Au at $\sqrt{s_{NN}}=$3 GeV are investigated within the UrQMD model. 
Simulations across a range of impact parameters $b$ reveal that the charged-pion yield ratio $\pi^{-}/\pi^{+}$ decreases with increasing $f_{n}$ but increases with $b$, exhibiting a pronounced dependence on both parameters, particularly at large impact parameters. 
Further, the two-pion correlation functions and corresponding HBT radii parameters for $\pi^{+}\pi^{+}$ and $\pi^{-}\pi^{-}$ are analysed with the CRAB program. 
The results show a clear charge splitting in both the correlation functions and the extracted HBT radii parameters, which becomes more prominent with increasing $f_{n}$ but remains only weakly sensitive to variations in $b$. 
These results suggest a potential new method to probe the neutron density distribution and the neutron skin thickness of nuclei, and infer the nuclear EoS through the charge splitting in the correlation function between $\pi^{+}\pi^{+}$ and $\pi^{-}\pi^{-}$ in HICs. 
Despite the debate on the observed charge splitting, further analyses with other transport models are appealing to ascertain the present results. 

In the present work, we primarily focus on the influence of the initial neutron density distribution on the final state observables related to charged pions, particularly the identical charged pion correlations. 
It is important to note, however, that in HICs at GeV energies, these observables can also be affected by other factors, such as the determination of the impact parameter, the experimental acceptance windows, and the understanding of high-density EoS and symmetry energy, as well as the third-body Coulomb effects at the final state. 
To get a more quantitative understanding of these correlations, especially the charge splitting between $\pi^{+}\pi^{+}$ and $\pi^{-}\pi^{-}$, further improving the consistency and predictive power of transport descriptions of HICs and a more rigorous comparison between experimental data and simulated results are needed. Given these complexities, more comprehensive and systematic analyses, such as those incorporating Bayesian inference, are warranted to constrain model parameters better and disentangle competing effects. 

\section*{Acknowledgements}
This work is supported in part by the National Natural Science Foundation of China under Grant (Nos. 12335008 and 12505143), 
the National Key Research and Development Program of China (Nos. 2023YFA1606402 and 2022YFE0103400), 
the Natural Science Foundation of Zhejiang Province (No. LQN25A050003),
the Natural Science Foundation of Huzhou City (No. 2024YZ28),
the Found of National Key Laboratory of Plasma Physics (No. 6142A04230203), 
the Scientific Research Fund of Zhejiang Provincial Education Department (No. Y202353782). 
The authors are grateful to the C3S2 computing center in Huzhou University and high performance computing center in  Tsinghua University for calculation support.

\end{document}